\documentclass[twocolumn]{aastex631}

\title{TOI-561 b}

\newcommand\aastex{AAS\TeX}

\usepackage{xcolor}
\usepackage{wrapfig}
\usepackage{url}
\usepackage{multirow}

\usepackage[T1]{fontenc} 
\usepackage[utf8]{inputenc} 

\usepackage{tipa}

\newcommand{\mb}{2.24} 
\newcommand{\mberr}{0.20} 

\newcommand{\mc}{6.6} 
\newcommand{\mcerr}{0.73} 

\newcommand{\md}{12.15} 
\newcommand{\mderr}{1.10} 

\newcommand{\me}{13.6} 
\newcommand{\meerr}{1.4} 

\newcommand{\radb}{1.37} 
\newcommand{\radberr}{0.04} 

\newcommand{\radc}{ 2.91} 
\newcommand{\radcerr}{0.04} 

\newcommand{\radd}{2.82} 
\newcommand{\radderr}{0.07} 

\newcommand{\rade}{2.55} 
\newcommand{\radeerr}{0.13} 

\newcommand{\kb}{2.18 } 
\newcommand{\kberr}{0.20} 

\newcommand{\db}{4.8} 
\newcommand{\dberr}{0.5} 

\received{\today}
\revised{TBD, 2018}
\accepted{TBD, 2018}

\shorttitle{\aastex\ TOI-561 b}
\shortauthors{Brinkman et al.}

\begin{document}

\title{TOI-561 b: A Low Density Ultra-Short Period ``Rocky'' Planet around a Metal-Poor Star}

\correspondingauthor{Casey L. Brinkman}
\email{clbrinkm@hawaii.edu}

\author[0000-0002-4480-310X]{Casey L. Brinkman}
\affiliation{Institute for Astronomy, University of Hawai'i, 2680 Woodlawn Drive, Honolulu, HI 96822 USA}

\author[0000-0002-3725-3058]{Lauren M. Weiss}
\affiliation{Department of Physics and Astronomy, University of Notre Dame, Notre Dame, IN, 46556, USA}

\author[0000-0002-8958-0683]{Fei Dai}
\affiliation{Division of Geological and Planetary Sciences, 1200 E California Blvd, Pasadena, CA, 91125, USA}

\author[0000-0001-8832-4488]{Daniel Huber}
\affiliation{Institute for Astronomy, University of Hawai'i, 2680 Woodlawn Drive, Honolulu, HI 96822 USA}

\author[0000-0002-1426-1186]{Edwin S. Kite}
\affiliation{Department of the Geophysical Sciences, University of Chicago, Chicago, IL}

\author[0000-0003-3993-4030]{Diana Valencia}
\affiliation{Centre for Planetary Sciences, University of Toronto, 1265 Military Trail, Toronto, ON, M1C 1A4, Canada}

\author[0000-0003-4733-6532]{Jacob L. Bean}
\affiliation{Department of Astronomy $\&$ Astrophysics, University of Chicago, 5640 S. Ellis Avenue, Chicago, IL 60637, USA}

\author[0000-0001-7708-2364]{Corey Beard}
\affiliation{Department of Physics $\&$ Astronomy, University of California Irvine, Irvine, CA 92697, USA}

\author[0000-0003-0012-9093]{Aida Behmard}
\affiliation{Department of Astronomy, California Institute of Technology, Pasadena, CA 91125, USA}

\author{Sarah Blunt}
\affiliation{Department of Astronomy, California Institute of Technology, Pasadena, CA 91125, USA}

\author[0000-0003-2404-2427]{Madison Brady}
\affiliation{Department of Astronomy $\&$ Astrophysics, University of Chicago, 5640 S. Ellis Avenue, Chicago, IL 60637, USA}

\author[0000-0003-3504-5316]{Benjamin Fulton}
\affiliation{NASA Exoplanet Science Institute/Caltech-IPAC, MC 314-6, 1200 E California Blvd, Pasadena, CA 91125, USA}

\author[0000-0002-8965-3969]{Steven Giacalone}
\affiliation{Department of Astronomy, 501 Campbell Hall, University of California, Berkeley, CA 94720, USA}

\author[0000-0001-8638-0320]{Andrew W. Howard}
\affiliation{Department of Astronomy, California Institute of Technology, Pasadena, CA 91125, USA}

\author[0000-0002-0531-1073]{Howard Isaacson}
\affiliation{Department of Astronomy, 501 Campbell Hall, University of California, Berkeley, CA 94720, USA}
\affiliation{Centre for Astrophysics, University of Southern Queensland, Toowoomba, QLD, Australia}

\author[0000-0003-0534-6388]{David Kasper}
\affiliation{Department of Astronomy $\&$ Astrophysics, University of Chicago, 5640 S. Ellis Avenue, Chicago, IL 60637, USA}

\author[0000-0001-8342-7736]{Jack Lubin}
\affiliation{Department of Physics $\&$ Astronomy, University of California Irvine, Irvine, CA 92697, USA}

\author[0000-0003-2562-9043]{Mason MacDougall}
\affiliation{Astronomy Department, 475 Portola Plaza, University of California, Los Angeles, CA 90095, USA}

\author[0000-0001-8898-8284]{Joseph M. Akana Murphy}
\affiliation{Department of Astronomy and Astrophysics, University of California, Santa Cruz, CA 95064, USA}

\author[0000-0002-9479-2744]{Mykhaylo Plotnykov}
\affiliation{Centre for Planetary Sciences, University of Toronto, 1265 Military Trail, Toronto, ON, M1C 1A4, Canada}

\author[0000-0001-7047-8681]{Alex S. Polanski}
\affiliation{Department of Physics and Astronomy, University of Kansas, Lawrence, KS, USA}

\author[0000-0002-7670-670X]{Malena Rice}
\altaffiliation{51 Pegasi b Fellow}
\affiliation{Department of Physics and Kavli Institute for Astrophysics and Space Research, Massachusetts Institute of Technology, Cambridge, MA 02139, USA}
\affiliation{Department of Astronomy, Yale University, New Haven, CT 06511, USA}

\author[0000-0003-4526-3747]{Andreas Seifahrt}
\affiliation{Department of Astronomy $\&$ Astrophysics, University of Chicago, 5640 S. Ellis Avenue, Chicago, IL 60637, USA}

\author[0000-0001-7409-5688]{Gu\dh mundur Stef\'ansson}
\affiliation{Department of Astrophysical Sciences, Princeton
University, 4 Ivy Lane, Princeton, NJ 08540, USA}
\affil{Henry Norris Russell Fellow}

\author{Julian St\"urmer}
\affiliation{Landessternwarte, Zentrum f\"ur Astronomie der Universit\"at Heidelberg, K\"onigstuhl 12, D-69117 Heidelberg, Germany}

\begin{abstract}

TOI-561 is a galactic thick disk star hosting an ultra-short period (0.45 day orbit) planet with a radius of 1.37 R$_{\oplus}$, making it one of the most metal-poor ([Fe/H] = -0.41) and oldest ($\approx$10 Gyr) sites where an Earth-sized planet has been found. We present new simultaneous radial velocity measurements (RVs) from Gemini-N/MAROON-X and Keck/HIRES, which we combined with literature RVs to derive a mass of M$_{b}$~=~\mb ~$\pm$ \mberr ~M$_{\oplus}$. We also used two new Sectors of TESS photometry to improve the radius determination, finding R$_{b}$~=~\radb ~$\pm$ \radberr ~R$_{\oplus}$, and confirming that TOI-561 b is one of the lowest-density super-Earths measured to date ($\rho_b$= \db ~$\pm$ \dberr ~g/cm$^{3}$). This density is consistent with an iron-poor rocky composition reflective of the host star's iron and rock-building element abundances; however, it is also consistent with a low-density planet with a volatile envelope. The equilibrium temperature of the planet ($\sim$2300 K) suggests that this envelope would likely be composed of high mean molecular weight species, such as water vapor, carbon dioxide, or silicate vapor, and is likely not primordial. We also demonstrate that the composition determination is sensitive to the choice of stellar parameters, and that further measurements are needed to determine if TOI-561 b is a bare rocky planet, a rocky planet with an optically thin atmosphere, or a rare example of a non-primordial envelope on a planet with a radius smaller than 1.5 R$_{\oplus}$.

\end{abstract}


\section{Introduction} 
\label{sec:intro}

TOI-561 b is a rare ultra-short period (USP, P=0.45 days) planet orbiting a star which originated in the galactic thick disk \citep{2021AJ....161...56W, 2021MNRAS.501.4148L}. It was the first chemically and kinematically confirmed thick-disk exoplanetary system detected by TESS, and is the first USP discovered around a thick disk star \citep{2021AJ....161...56W}. At 1.4 times the radius of the Earth, TOI-561 b is a super-Earth size planet, and the TOI-561 system is one of only two confirmed planetary systems discovered around thick disk stars to date (the other is Kepler-444, \cite{2015ApJ...799..170C})\footnote{Additional planets have been validated around thick-disk stars, but not confirmed with RV mass measurements, such as LHS 1518b \citep{2020AJ....159..160G}}. TOI-561 is 10 $\pm$ 3 Gyr old \citep{2021AJ....161...56W}, making TOI-561 b one of the oldest known super-Earth sized planets, indicating that rocky planets have been forming for nearly the age of the universe, even in metal-poor environments. Previous studies of this system suggest an even more unique aspect of this planet: it is the lowest-density super-Earth discovered to date, and potentially inconsistent with having a rocky planet composition \citep{2021MNRAS.501.4148L, 2022MNRAS.511.4551L}. 

Two groups have previously published mass and radius measurements for TOI-561 b, along with the three additional sub-Neptune sized planets in the system. \cite{2021AJ....161...56W} used RVs from the Keck/HIRES spectrograph and found a mass of M$_{b}$=3.24 $\pm$ 0.83 M$_{\oplus}$, a radius of R$_{b}$=1.45 $\pm$ 0.11 R$_{\oplus}$, and a density of $\rho_{b}$=5.6 $\pm$ 2.2 g/cm$^3$. This is approximately one standard deviation less dense than a planet with an Earth-like composition would be at R$_{b}$=1.45 R$_{\oplus}$. \cite{2021MNRAS.501.4148L} used RVs from the TNG/HARPS-N spectrograph and found an even lower mass of M$_{b}$=1.59 $\pm$ 0.36 M$_{\oplus}$, a radius of R$_{b}$=1.43 $\pm$ 0.11 R$_{\oplus}$, with a corresponding density of $\rho_{b}$=3.0 $\pm$ 0.8  g/cm$^3$. More recently, \cite{2022MNRAS.511.4551L} combined the literature RVs from HIRES and HARPS-N with new HARPS-N RVs, finding a mass of M$_{b}$=2.00 $\pm$ 0.23 and a density of $\rho_{b}$=3.8 $\pm$ 0.5 g/cm$^3$. Based on their analysis, TOI-561 b is inconsistent with a rocky composition and was suggested to have a water steam envelope.

From the solar system, we expect large planets to have extensive low molecular weight envelopes, while smaller planets are composed primarily of rock and metal. The masses and radii of small exoplanets suggest a transition between primarily rocky and gas-enveloped planets at approximately 1.5 $R_{\oplus}$ \citep{2014ApJ...783L...6W, 2015ApJ...801...41R, 2017AJ....154..109F}, with planets smaller than 1.5 $R_{\oplus}$ often having compositions consistent with Earth-like iron-to-silicate ratios \citep{2015ApJ...800..135D}. However, existing super-Earth mass measurements indicate a wide diversity of densities among those planets with R$>$1.5 $R_{\oplus}$---far more diverse than we observe for rocky planets in our own solar system \citep{2014ApJS..210...20M, 2016ApJ...822...86M, 2019ApJ...883...79D}. These densities suggest the interior compositions of Earth and super-Earth sized planets could potentially vary from entirely made of silicate rock, to predominantly made of iron \citep{2019NatAs...3..416B}, with high-molecular-mass atmospheres possible \citep{2017AJ....154..232A, 2021ApJ...909L..22K}.

Because exoplanets are born from the same primordial nebular material as their host star, the abundances of refractory elements in planets should correlate with those of their host star, and stellar abundances can be used as a prior for rocky planet chemical composition \citep{2015A&A...577A..83D}. Comparing the population of rocky exoplanets to the population of host stars, we see that planets span a wider range in refractory abundances than stars \citep{2020MNRAS.499..932P}. When looking at one-to-one comparisons of rocky exoplanets and their host stars, the error bars--especially in mass--are too large in most cases to draw definite conclusions \citep{2020MNRAS.499..932P, 2021PSJ.....2..113S}, except for a few cases where planets differ in composition by one sigma compared to their star \citep{2021PSJ.....2..113S}. However, when performing one-to-one comparisons there appears to be a correlation between stellar and planet enrichment \citep{2021Sci...374..330A}. Improving the mass and radius estimates on TOI-561 b provides a unique opportunity to compare the planetary and stellar compositional similarities for such an old, highly-irradiated planet.
  
In this paper, we further refine the radius and mass measurements for TOI-561 b, along with our understanding of its bulk composition. We first use two additional Sectors of TESS photometry to measure the radius of the planet. We then present the results of the first simultaneous radial velocity program combining MAROON-X \citep{2018SPIE10702E..6DS, 2020SPIE11447E..1FS}, a new fiber-fed RV spectrometer on Gemini-N, and HIRES, a well-characterized spectrometer on Keck I to measure the mass of the planet in combination with literature RVs. We then combine these mass and radius measurements to investigate the potential compositions of TOI-561 b, and the sensitivity of these results to stellar parameter choice and assumptions about the mantle melt fraction.

\section{Planet Radii}
The TESS spacecraft has observed TOI-561 in Sectors 8, 35, 45, and 46. \cite{2021AJ....161...56W} and \cite{2021MNRAS.501.4148L} used Sector 8, while \cite{2022MNRAS.511.4551L} used Sectors 8 and 35. We used all four available Sectors, analyzing both TESS's typical 2-minute cadence and 20-sec cadence light curves for Sectors 45 and 46. The addition of 20-second data is particularly valuable, since it has shown to yield improved photometric precision for bright stars \citep{2022AJ....163...79H}. 

To improve the constraints on the planet radii, we downloaded the Simple Aperture Photometry (SAP) light curves of relevant TESS Sectors from the Mikulski Archive for Space Telescopes (MAST) at the Space Telescope Science Institute. The specific observations analyzed can be accessed via \dataset[DOI]{http://dx.doi.org/10.17909/7y1e-1k46}. Sector 8 photometry featured two transits that \cite{2021AJ....161...56W} interpreted as two transits of the same planet at 16.287 $\pm$ 0.005 days (giving a total of 3 transiting planets for the system), while \cite{2021MNRAS.501.4148L} interpreted it as single transits of two different planets (giving 4 transiting planets). Follow up photometry from CHEOPS lifts this degeneracy in favor of the four planet model with planet d at an orbital period of 25.7124 $\pm$ 0.0002 days and planet e at a period of 77.03 $\pm$ 0.25 days \citep{2022MNRAS.511.4551L}. The additional TESS observations of Sectors 45 and 46  further support the four planet model as described in \cite{2022MNRAS.511.4551L}. 

Our analyses employs the Python package {\tt Batman} \citep{Kreidberg2015} to model the transit light curves. We imposed a Gaussian prior on the host star mean density of $1.38\pm0.11\rho_\odot$ \citep{2021AJ....161...56W}. The mean stellar density is a global parameter that is used to generate the transit light curves of all planets in the TOI-561 system consistently \citep{Seager}. Two other global parameters are the quadratic limb darkening coefficients $q1$ and $q2$ as parameterized by \citet{Kipping}. Assuming circular orbits, each planet also has the following transit parameters: the orbital period $P_{\text{orb}}$, the time of conjunction $T_{\text{c}}$, the planet-to-star radius ratio $R_{\text{p}}/R_\star$, and the impact parameter $b\equiv a\cos i/R_\star$. The semi-major axis in units of stellar radii $a/R_\star$ is implicitly constrained based on the orbital periods and the host star density. 

For each planet, we started from the transit parameters reported by the TESS team on ExoFOP \footnote{\url{https://exofop.ipac.caltech.edu}.}. We isolated the individual transit with a window size of 2 times the reported transit duration.  After removing any overlapping transits, we analyzed 193, 8, 4 and 2 individual transits analyzed for planets b, c, d and e, respectively. In our final global fit, we analyzed all transits simultaneously. We treated any overlapping transits as a simple sum of the individual transit light curves (i.e. ignoring any possible planet-planet eclipse). We fitted and removed any local stellar variability with a quadratic polynomial. We then fit all transits of each planet together with the Levenberg-Marquardt optimization in {\tt Python} package {\tt lmfit} \citep{newville_matthew_2014_11813}. The best-fit model then served as a new template transit model to fit the individual transit and local stellar variability. We repeated this process three times. We did not detect any quasi-sinusoidal variation in the individual transit times that would hint at transit timing variations.

Finally, we sampled the posterior distribution of a global transit model with the affine-invariant Markov Chain Monte Carlo method using the {\tt Python} package {\tt emcee} \citep{2013PASP..125..306F}. We initialized 128 walkers near the best-fit Levenberg-Marquardt solution found by {\tt lmfit}. We ran the code for 50000 links after which we checked for convergence by calculating the autocorrelation lengths for each parameter which range from 30-200 samples. For even the slowest converging parameter (200 samples), we are using more than 250x more samples than the autocorrelation length, making convergence a safe assumption\footnote{The suggested sample size to ensure convergence is N > 50$\tau$ where $\tau$ is the autocorrelation length.}\citep{emcee}.

The orbital periods and times of conjunction of the planets are summarized in Table \ref{table:radvel}. Our transit-based ephemeris were used to further model the radial velocity measurements (\S\ref{sec:observations}). Crucially, the planet/star radius ratio of planet b was constrained to be 0.01507$\pm$0.00043. Using the stellar radius from \cite{2021AJ....161...56W}, this translates to a planet radius of R$_{b}$=\radb $\pm$ \radberr.

We additionally analyzed the TESS PDCSAP data to look for any potential variations in the transit depth between TESS Sectors using the {\tt exoplanet} package\citep{exoplanet:zenodo} and its dependencies \citep{exoplanet:agol20,
exoplanet:arviz, exoplanet:astropy13, exoplanet:astropy18, exoplanet:kipping13,
exoplanet:luger18, exoplanet:pymc3, exoplanet:theano}. We downloaded the \textit{TESS} Pre-search Data Conditioning SAP (PDCSAP) flux photometry using {\tt Lightkurve} \citep{2018ascl.soft12013L}, and masked the transits of planets c, d, and e. We then imposed a Gaussian prior on stellar density (same as above), and used Gaussian priors on period and conjunction time for TOI-561 b using the values and uncertainties reported in \cite{2022MNRAS.511.4551L}. We then created a transit model using limb darkening parameters from \cite{exoplanet:kipping13} (also described above). We then ran a Hamiltonian Monte Carlo using {\tt PyMC3} \citep{Salvatier2016} to sample the posterior distribution using 20 chains, 10000 draws, and checked for convergence by ensuring the number of draws is $>$50 times greater than the longest autocorrelation length.

We fit each Sector of data, and each available cadence, individually and combined to see the variation in transit depth between Sectors and observation cadence (Table \ref{table:sectors}). The Sectors with the largest difference in R$_{P}$/R$_{*}$ (Sector 8 2-min cadence, and Sector 46 20-second cadence) differ by 2$\sigma$. The combined fits from all Sectors and cadences agrees with the R$_{P}$/R$_{*}$ found using {\tt Batman}. The values for transit depth reported are from {\tt Batman}.

\begin{table}[]
\begin{center}
\begin{tabular}{|l|l|}
\hline
Transit Parameter & Posterior Results\\       
\hline
$\rho_\star$ ($\rho_\odot$) & 1.29$\pm$ 0.04\\
\hline
$q_1$  & 0.34$\pm$ 0.20\\
\hline
$q_2$  & 0.32$\pm$ 0.22\\
\hline
$r_p/r_\star$  & 0.01507$\pm$ 0.00043\\
\hline
$i_{\rm orb} (^{\circ})$  & 87.9$\pm$ 1.8\\
\hline
$a/R_\star$  & 2.669$\pm$ 0.034\\

\hline
$b$  & 0.10$\pm$0.08\\
\hline
$e$  & 0 (fixed)\\
\hline
$\omega$  & 0 (fixed)\\
\hline
$P_{\rm orb}$ (days)  & 0.4465690$\pm$0.0000012 \\
\hline
$T_c$ (BJD-2457000)  & 1517.4984 $\pm$ 0.0019\\
\hline
\end{tabular}
\caption{Transit Parameters of TOI-561 b based on Sectors 8 and 35 (2 minute cadence) as well as 45 and 46 (both 2 minute and 20 second cadence available). Values were found using {\tt Batman}.}
\label{table:transit}
\end{center}
\end{table}

\begin{figure}
    \centering
    \includegraphics[width=0.5\textwidth]{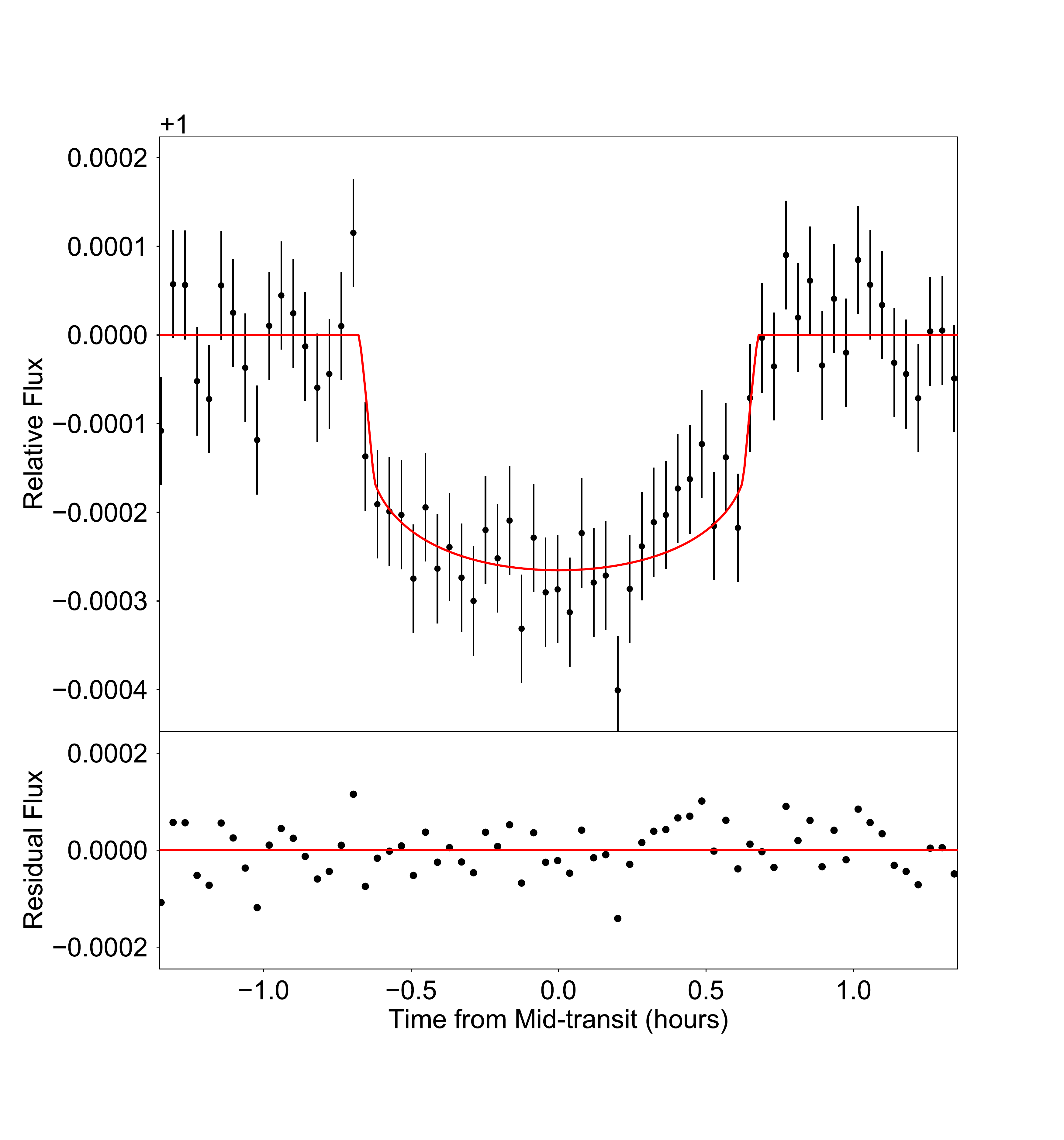}
    \caption{Phase-folded $\sc TESS$ transit light curve of TOI-561 b from all available Sectors. We have removed overlapping transits with other planets. We have binned the phased light curves (black)}. The red curve shows the best-fit transit model.
    \label{fig:transit}
\end{figure}

\begin{table*}[]
\begin{center}
\begin{tabular}{|l|l|l|l|l|l|}
\hline
Sector    & 8                 & 35                & 45                & 46                 & Joint-fit        \\  \hline
2 Minute  & 0.0155$\pm$0.0009 & 0.0149$\pm$0.0011 & 0.0141$\pm$0.0009 & 0.01460$\pm$0.0009 & 0.0155$\pm$0.0005\\ \hline
20 Second &    N/A            &    N/A            & 0.0147$\pm$0.0008 & 0.01396$\pm$0.0009 & 0.0145$\pm$0.0007 \\
\hline
\end{tabular}
\caption{R$_{P}$/R$_{*}$ measured for each available Sector (8, 35, 45, 46) of TESS photometry, and where available with both 2 Minute and 20 Second cadence data.}
\label{table:sectors}
\end{center}
\end{table*}

\section{Planet Masses} 
\label{sec:observations}

\begin{table}[]
\begin{center}

\begin{tabular}{|l|l|l|l|}
\hline
Time          & RV         & $\sigma_{RV}$ & Spectrograph  \\ 
(BJD)         & (m/s)      & (m/s)         &   \\ \hline
 2459267.750  & -3.7  & 1.6      & MAROON-X Blue   \\ \hline
 2459267.813  & -5.3  & 1.4      & MAROON-X Blue   \\ \hline
 2459267.750  & -6.54  & 2.9      & MAROON-X Red    \\ \hline
 2459267.813  & -12.4 & 2.7      & MAROON-X Red    \\ \hline
 2459632.851  & -1.9  & 1.19      & HIRES   \\ \hline
 2459649.771  & 0.1   & 1.5      & HIRES   \\ \hline
\end{tabular}
\caption{A Few lines of our RV Table are shown here, but the full machine-readable table will be available online.}
\label{table:rvs}
\end{center}
\end{table}

\subsection{HIRES RVs}

Our analysis incorporates 102 total RVs from the High Resolution Echelle Spectrograph (HIRES) on the W. M. Keck Observatory 10m telescope Keck-I on Maunakea, Hawai`i \citep{1994SPIE.2198..362V}. \cite{2021AJ....161...56W} obtained 60 RVs between May 2019 and October 2020 through the TESS-Keck Survey (TKS) collaboration \citep{2021arXiv210606156C}. We collected 42 additional RVs from October 2020-March 2022.\footnote{Telescope time was allocated by University of Hawaii, University of California, California Institute of Technology, and NASA.}  The RVs used in this paper are listed in Table \ref{table:rvs}. 

We used the standard California Planet Search (CPS) data reduction pipeline as described in \cite{2010ApJ...721.1467H}. This method uses an iodine cell mounted in front of the slit in order to provide a provide a wavelength reference \citep{1992PASP..104..270M}.  Sky subtraction was performed as part of the raw reduction through the use of a $14\arcsec.0$ long slit in order to spatially resolve the sky with respect to the seeing-limited point-spread function (full-width half-max $\approx$ $1\arcsec.0$). Measuring the RVs requires characterizing the PSF of the spectrometer, which is time-variable due primarily to changing seeing and weather.  The CPS Doppler routine involves forward-modeling the iodine-imprinted spectrum of a star as the combination of a library iodine spectrum and a velocity-shifted, iodine-free, PSF-deconvolved template spectrum of the target star, the combination of which is then convolved with the best-fit PSF. To deconvolve the PSF from the iodine-free template, we observed rapidly-rotating B stars with the iodine cell in the light path immediately before and after the template, effectively sampling the PSF at the time of the template in the iodine absorption profiles.

\subsection{HARPS-North RVs}
Our analysis also incorporates 143 published RVs from the HARPS-N spectrograph installed on the 3.6 meter Telescopio Nazionale Galileo (TNG) at the Observatorio Roque de Los Muchachos in La Palma, Spain. Originally published in \cite{2021MNRAS.501.4148L} and \cite{2022MNRAS.511.4551L}, these RVs were collected between November 2019 and June 2021. 

\subsection{MAROON-X RVs}
MAROON-X is a new, state-of-the-art fiber-fed spectrograph mounted on the 8.1 meter Gemini-North telescope on Maunakea, Hawai`i. It operates in the red-optical (500-920nm) with resolving power R$\approx$85,000, and uses both red and blue arms to get two radial velocity measurements per exposure \citep{2016SPIE.9908E..18S, 2018SPIE10702E..6DS, 2020SPIE11447E..1FS}. MAROON-X has demonstrated a stability of 30 cm/s, and has been used to measure some of the most precise masses for rocky planets in the literature to date \citep{2021Sci...371.1038T, 2022AJ....163..168W}. 

Our observations used the simultaneous calibration fiber of MAROON-X, which allows for a robust order-by-order drift correction to sub-m/s precision. The raw data was reduced using a custom pipeline based on that developed for CRIRES \citep{2010Msngr.140...41B}, and RVs were computed using \texttt{SERVAL} \citep{2018A&A...609A..12Z}. A full description of MAROON-X data reduction can be found in \cite{2022AJ....163..168W}.

We collected a total of 70 RVs between February 2021 and May 2021, 35 individual exposures with RVs from the Red and Blue arms\footnote{Telescope time was allocated by the University of Hawaii.}. We used an integration time of 460 seconds, and we found that the blue arm RVs had a median uncertainty of 1.4 m/s, while those from the red arm had a median uncertainty of 2.8 m/s.

\begin{figure*}
    \centering
    \includegraphics[width=1.0\textwidth]{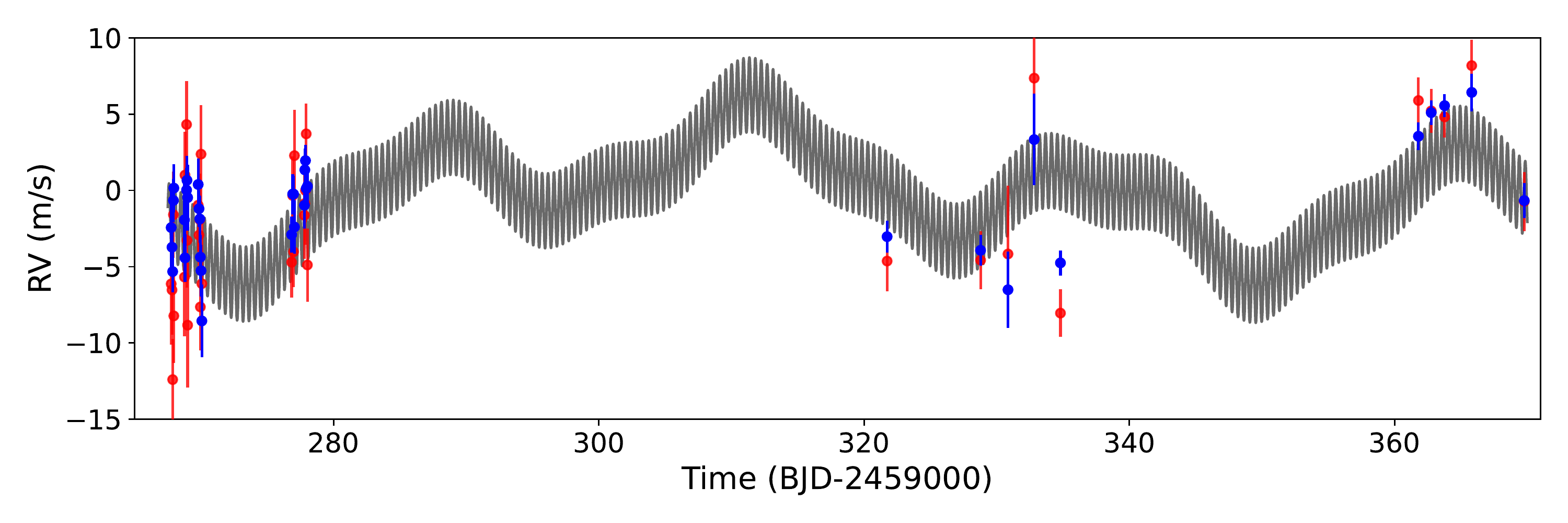}
    \includegraphics[width =1.0\textwidth]{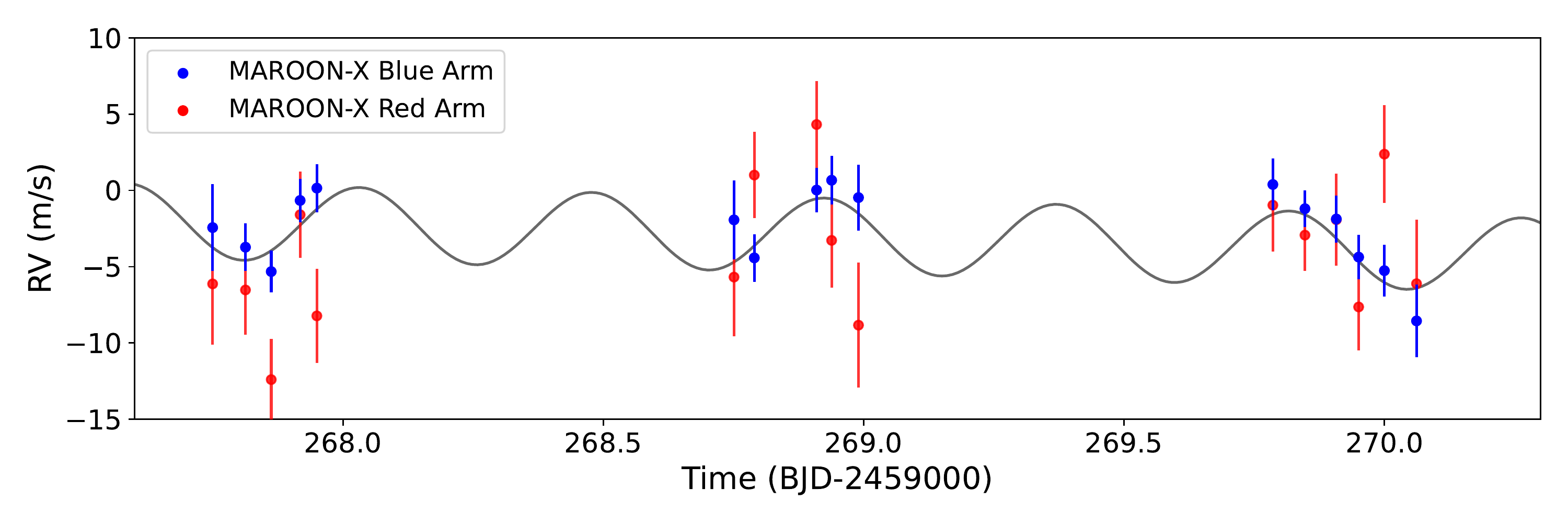}

    \caption{Top: The MAROON-X RVs collected from February to May 2021 over three different runs (red-arm RVs shown in red, blue-arm RVs shown in blue). The grey line is the best-fit Keplerian orbital model for all four planets in the TOI-561 system. Bottom: RVs from our highest-cadence run in February 2021, highlighting the fully sampled orbit of TOI-561 b.}
    \label{fig:maroonx}
\end{figure*}

We observed TOI-561 with a mixture of high-cadence runs and long term monitoring. High-cadence observations (5+ RVs per night) allowed us sample the full phase curve of 0.45 day planet TOI-561 b, while freezing out the contribution of longer period planets. During our highest cadence run, we collected 5-6 RVs per night for three nights in February 2021 (Figure \ref{fig:maroonx}). We also collected simultaneous observations on HIRES during this run to verify the stability of MAROON-X. 

\begin{figure*}
    \centering
    \includegraphics[width=1.0\textwidth]{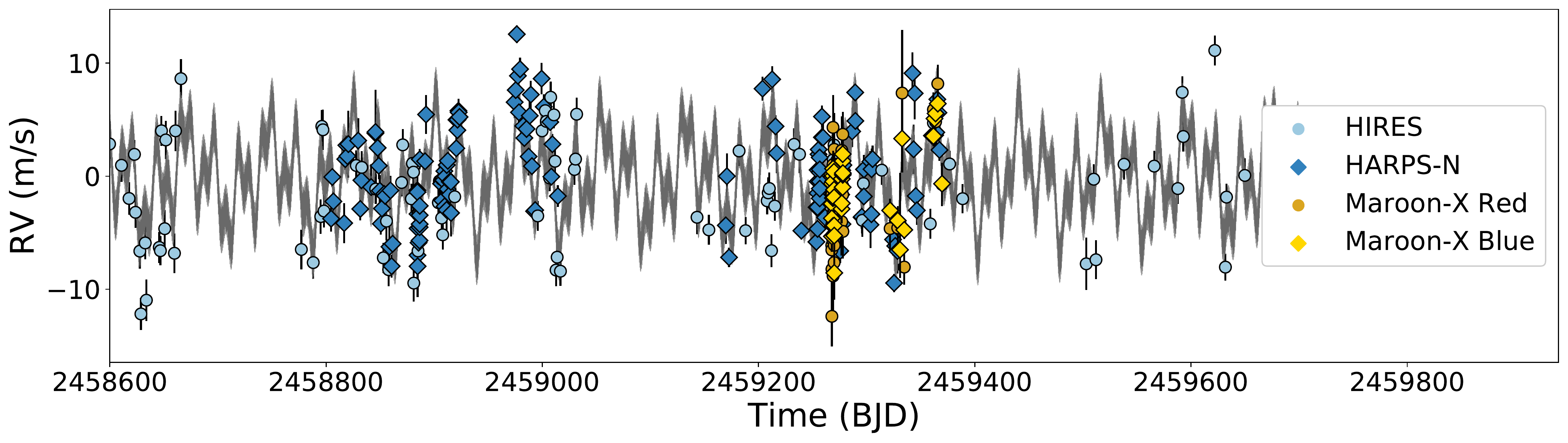}
    \caption{Radial Velocity vs. time of TOI-561 measured with HIRES, HARPS-North, and MAROON-X (both red and blue arms), with 1$\sigma$ error bars. The grey line represents our best-fit four-planet Keplerian orbital model.}
    \label{fig:rvs}
\end{figure*}

\subsection{Keplerian Orbital Fit}

\label{sec:radvel}
We used the open source python package \texttt{RadVel} \citep{2018PASP..130d4504F} to model the RVs. We measured the mass of each planet by fitting the RVs for a Keplerian orbit, in which the RV curve is described by the orbital period (P), conjunction time (T$_{c}$), eccentricity (e), argument of periastron, and RV semi-amplitude (K) of each planet. We include two additional terms per dataset to fit the RVs: a zeropoint offset ($\gamma$) and an RV jitter term ($\sigma_{j}$). Jitter accounts for additional Gaussian noise that can be astrophysical in origin, or can come from systematics of the spectrograph. This additional uncertainty was added in quadrature with the intrinsic uncertainties on the RVs during our optimization of the likelihood function and Markov Chain Monte Carlo analysis (described below).

Our constraints on period and conjunction time for TOI-561 b and c from TESS photometry are more precise than we would be able to measure using RVs, therefore we fix these values to the photometric ones in the RV fit. As a test, we also fit the RVs with gaussian priors on conjunction time and period centered on the best-fit values from TESS photometry, and we recover posterior values consistent with the solution using fixed period and conjunction time. We assume that the USP has a circular orbit \citep{ 2021AJ....161...56W}, and we also assume low eccentricity orbits for the three outer planets, as is typical for compact multi-planet systems  \citep{2013ApJ...774L..15D, 2019AJ....157...61V, 2019AJ....157..198M, 2021AJ....162...55Y}. As a test, we used both models that allow the eccentricities to vary with no prior, and models with eccentricity fixed at zero. The best-fit solution in either model prefers circular orbits, and as a consequence of the eight additional parameters produces a larger BIC for the model that allows variable eccentricity(1630 vs 1580 with fixed circular orbits), therefore we adopt eccentricities fixed at zero. Our model parameter constraints and posterior values are summarized in Table \ref{table:radvel}. After optimizing for the maximum likelihood fit, we ran \texttt{RadVel}'s built-in Markov Chain Monte Carlo algorithm \citep{2013PASP..125..306F} to explore the surrounding parameter space and estimate the uncertainty in the model parameters, and to explore the covariance between parameters (no strong co-variances or degeneracies were found).

\begin{table*}[]
\begin{center}
    
\begin{tabular}{|c|c|c|c|c|c|}
\hline
 \multicolumn{2}{|c|}{}     & Parameter & Unit & Median Posterior Value &  Prior \\ \hline
 \multicolumn{2}{|c|}{\multirow{4}{*}{TOI-561 b}}  & P          & Days & 0.4465688 &  \multirow{3}{*}{Fixed}\\\cline{3-5} 
 \multicolumn{2}{|c|}{}     & T$_{c}$        & BJD  & 2458686.30 & \\\cline{3-5}
 \multicolumn{2}{|c|}{}     & e         &      &  0         &\\\cline{3-6}
 \multicolumn{2}{|c|}{}     & K         & m/s  & \kb $\pm$ \kberr  & Uniform \\\hline
 \multicolumn{2}{|c|}{\multirow{4}{*}{TOI-561c}} & P         & Days & 10.778831    & \multirow{3}{*}{Fixed} \\\cline{3-5} 
 \multicolumn{2}{|c|}{}     & T$_{c}$        & BJD  & 2458527.06 & \\\cline{3-5}
\multicolumn{2}{|c|}{}      & e         &      &  0         &\\\cline{3-6}
\multicolumn{2}{|c|}{}      & K         & m/s  & 2.22 $\pm$ 0.23 & Uniform \\\hline
\multicolumn{2}{|c|}{\multirow{4}{*}{TOI-561d}} & P         & Days & 25.7126    & \multirow{3}{*}{Fixed} \\\cline{3-5} 
\multicolumn{2}{|c|}{}      & T$_{c}$        & BJD  & 2458521.88 & \\\cline{3-5}
\multicolumn{2}{|c|}{}      & e         &      &  0         &\\\cline{3-6}
\multicolumn{2}{|c|}{}      & K         & m/s  & 3.04 $\pm$ 0.25  & Uniform \\\hline
\multicolumn{2}{|c|}{\multirow{4}{*}{TOI-561e}} & P         & Days & 77.1437    & \multirow{3}{*}{Fixed} \\\cline{3-5} 
\multicolumn{2}{|c|}{}      & T$_{c}$        & BJD  & 2458538.18 & \\\cline{3-5}
\multicolumn{2}{|c|}{}      & e         &      &  0         &\\\cline{3-6}
\multicolumn{2}{|c|}{}      & K         & m/s  & 2.36 $\pm$ 0.23  & Uniform \\\hline
\multicolumn{2}{|c|}{\multirow{2}{*}{HIRES}} & $\sigma$  & m/s  & 2.8 $\pm$ 0.3  & Hardbound 0,10\\\cline{3-6}
\multicolumn{2}{|c|}{}      & $\gamma$  & m/s  & -1.7 $\pm$ 0.3  & Uniform \\\hline
\multicolumn{2}{|c|}{\multirow{2}{*}{HARPS-N}} & $\sigma$  & m/s  & 2.0 $\pm$ 0.2  & Hardbound 0,10\\\cline{3-6}
\multicolumn{2}{|c|}{}      & $\gamma$ & m/s  & -0.004  $\pm$ 0.20     & Uniform \\\hline
\multirow{6}{*}{MAROON-X} & \multirow{2}{*}{ February} & $\sigma$  & m/s  & 0.40 $\pm$ 0.3 & Hardbound 0,10\\\cline{3-6}
\multirow{6}{*}{Blue}     &  & $\gamma$  & m/s  & -2.4 $\pm$ 0.4 & Uniform \\\cline{2-6}
                          & \multirow{2}{*}{April} & $\sigma$  & m/s  & 1.6  $\pm$ 0.9      & Hardbound 0,10\\\cline{3-6}
                          & & $\gamma$  & m/s  & -1.7 $\pm$ 1.1 & Uniform \\\cline{2-6}
                          & \multirow{2}{*}{May} & $\sigma$  & m/s  & 1.7  $\pm$ 0.8 & Hardbound 0,10\\\cline{3-6}
                          & & $\gamma$  & m/s  & -1.8 $\pm$ 0.93   & Uniform \\\hline
\multirow{6}{*}{MAROON-X} & \multirow{2}{*}{February} & $\sigma$  & m/s  & 2.74 $\pm$ 0.95      & Hardbound 0,10\\\cline{3-6}
\multirow{6}{*}{Red}      & & $\gamma$  & m/s  & -3.63 $\pm$ 0.83  & Uniform \\\cline{2-6}
                          & \multirow{2}{*}{April} & $\sigma$  & m/s  & 3.67 $\pm$ 1.82      & Hardbound 0,10\\\cline{3-6}
                          & & $\gamma$  & m/s  & -2.50 $\pm$ 1.22     & Uniform \\\cline{2-6}
                          & \multirow{2}{*}{May} & $\sigma$  & m/s  & 0.44 $\pm$ 1.25      & Hardbound 0,10\\\cline{3-6}
                          & & $\gamma$  & m/s  & -1.06 $\pm$ 0.95     & Uniform 0,10\\\hline

\end{tabular}
\caption{\texttt{Radvel} model parameters for the Keplerian orbit of the TOI-561 system. The period and conjunction time are taken from Table \ref{table:transit}, and are more precise than we would be able to determine using RVs, therefore we fixed them in our analysis. Jitter ($\sigma$) and RV offset ($\gamma$) for each instrument are also listed, with unique jitter and offsets allowed for each run with MAROON-X.}
\label{table:radvel}
\end{center}

\end{table*}

 \begin{figure*}
    \centering
    \includegraphics[width=0.6\textwidth]{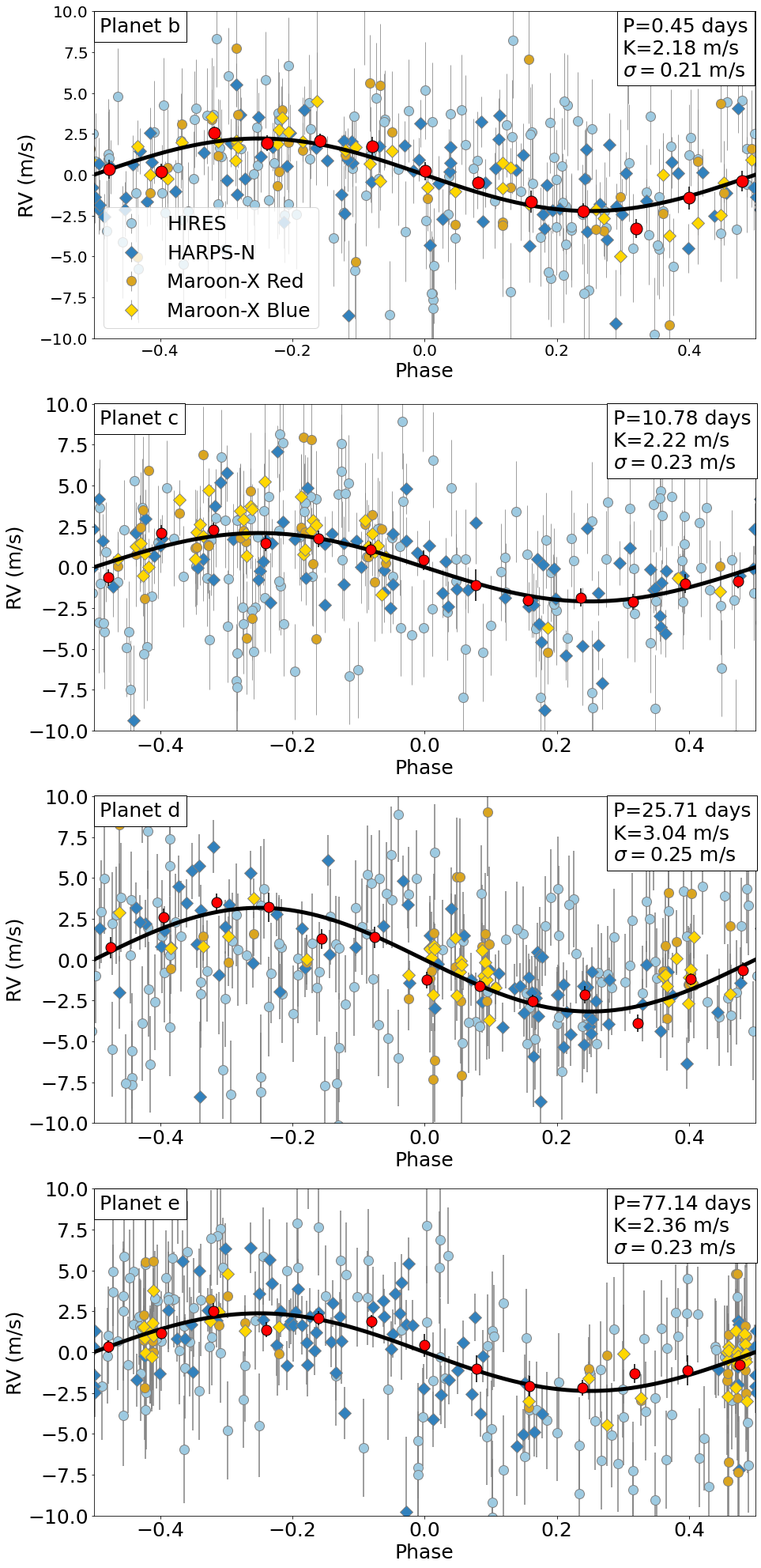}
   
    \caption{Top panel: the radial velocities phase-folded at the period of TOI-561 b, after subtracting the RV components from the other three planets based on our best-fit model. The model RV curve for planet b is overplotted in black, with the model period, semi-amplitude (K), and the standard deviation in semi-amplitude ($\sigma$) shown. The phase-folded weighted mean RVs and their uncertainties are shown in red.  Middle and bottom panels: same as the top panel, but for planets c, d, and e.}
    \label{fig:phasefold}
\end{figure*}

The phase-folded best-fit models for each planet are shown in Figure \ref{fig:phasefold}. We computed the mass for each planet using the best-fit semi-amplitudes, stellar mass from \cite{2021AJ....161...56W}, and planet orbital periods (Table \ref{table:radvel}). Using the masses and radii measured here we also computed the bulk density of each planet (Table \ref{table:properties}). 

\begin{table*}[]
\begin{center}

\begin{tabular}{|l|c|c|c|c|}
\hline
 Planet Name                     & b                        & c                      & d                     & e \\ \hline
 Period, $P$ (days)              & 0.4465688 $\pm$ 0.0000008 & 10.778831 $\pm$ 0.000036 & 25.7124 $\pm$ 0.0002  & 77.03 $\pm$ 0.25 \\ \hline
 Semi-Major Axis, $a$ (AU)       & 0.0106 $\pm$ 0.0004      & 0.089 $\pm$ 0.003      & 0.159 $\pm$ 0.006     & 0.33 $\pm$ 0.01 \\ \hline
 Radius, $R$ (R$_{\oplus}$)      & \radb $\pm$ \radberr     & \radc $\pm$ \radcerr   & \radd $\pm$ \radderr  & \rade $\pm$ \radeerr \\ \hline
 Semi-Amplitude, $K$ (m/s)       & \kb  $\pm$ \kberr        & 2.22 $\pm$ 0.23        & 3.04 $\pm$ 0.25       & 2.36 $\pm$ 0.23 \\ \hline
 Mass, $M$ (M$_{\oplus}$)        & \mb $\pm$ \mberr         & \mc $\pm$ \mcerr       & \md $\pm$ \mderr      & \me $\pm$ \meerr \\ \hline
 Density, $\rho$ (g/cm$^{3}$)    &  \db   $\pm$  \dberr     & 1.46 $\pm$  0.19       & 2.81 $\pm$ 0.33       & 4.91 $\pm$ 0.74 \\ \hline

\end{tabular}
\caption{Directly modeled and derived parameters based on our MCMC analysis of the four transiting planets in TOI-561. The stellar mass (0.805 M$_{\odot}$) and radius (0.832 R$_{\odot}$) used to compute planet mass and radii are isochrone values from \cite{2021AJ....161...56W}}
\label{table:properties}
\end{center}

\end{table*}

For the USP TOI-561 b, the posterior of our fit yields a mass of M$_{b}$~=~\mb~ $\pm$ \mberr ~M$_{\oplus}$. This is 1.2$\sigma$ below the previously published value from \cite{2021AJ....161...56W} using observations solely from HIRES, and 1.8$\sigma$ above the initial value published by \cite{2021MNRAS.501.4148L} using observations solely from HARPS-N. It is in agreement within 1$\sigma$ of the updated mass by \cite{2022MNRAS.511.4551L} which utilized both HIRES (60) and HARPS-N (144) RVs.

We also measured the mass of TOI-561 b using just the blue-arm RVs from MAROON-X during our February high-cadence run to highlight the unique abilities of MAROON-X. Using these 16 RVs, we measured a planet mass of M$_{b}$=2.7 $\pm$ 0.48 M$_{\oplus}$. MAROON-X was able to achieve a 17$\%$ fractional uncertainty with 16 RVs, compared to a fractional uncertainty of 35$\%$ using 80 measurements with HIRES. The value agrees with the mass derived using all three spectrographs, and is closer to the new best-fit mass than the initial published values using only HIRES or HARPS-N. The combination of HARPS-N, HIRES, and MAROON-X data allowed us to measure the mass of TOI-561 b with a fractional uncertainty of 9.6$\%$, placing it among the most precisely known RV masses for planets with R$_{p} <$1.5R$_{\oplus}$ (see \citealt{2019ApJ...883...79D, 2021Sci...371.1038T, 2021A&A...649A.144S} for other examples). 

\begin{figure*}
    \centering
    \includegraphics[width=1.0\textwidth]{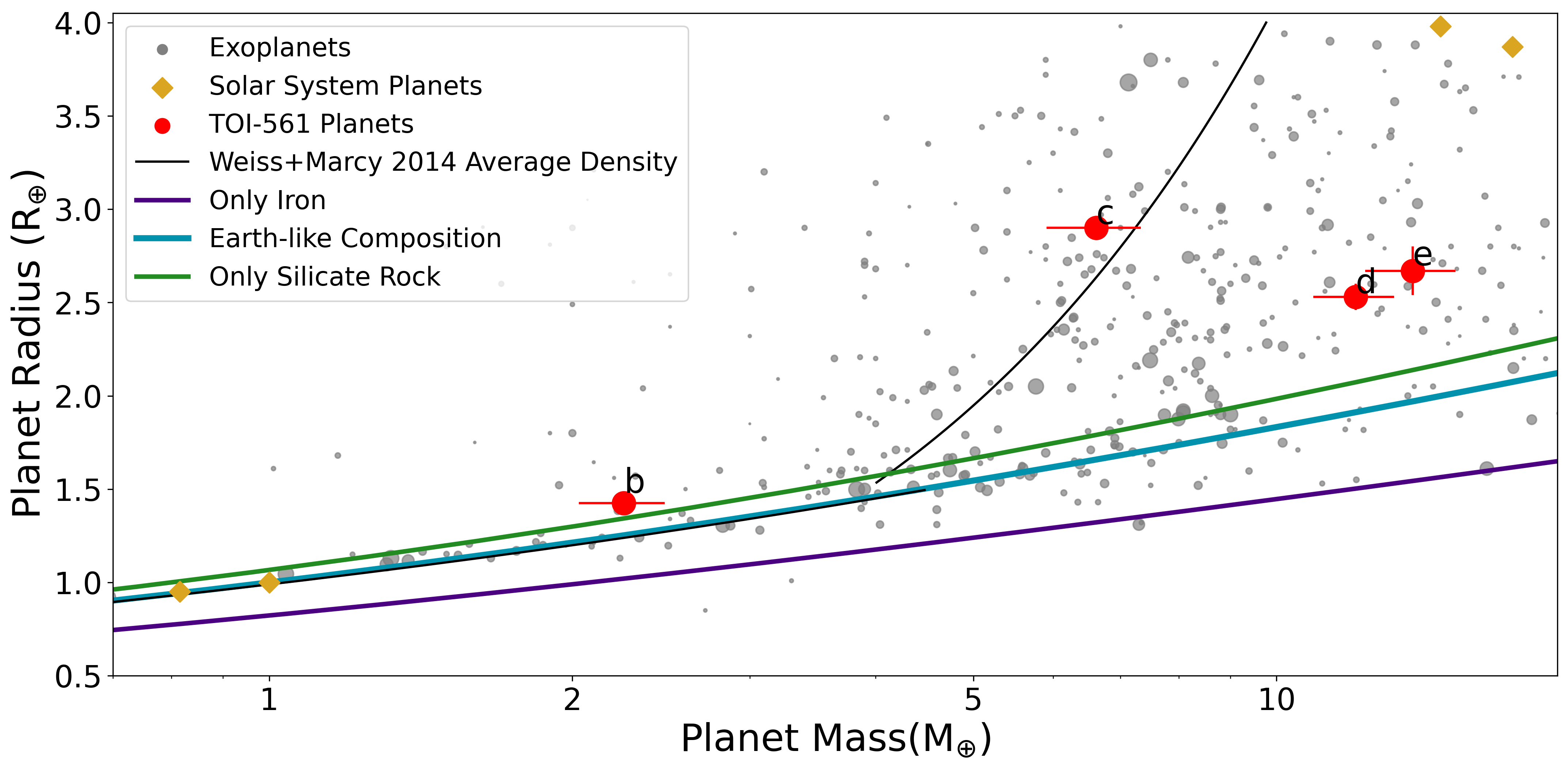}
    \includegraphics[width=1.0\textwidth]{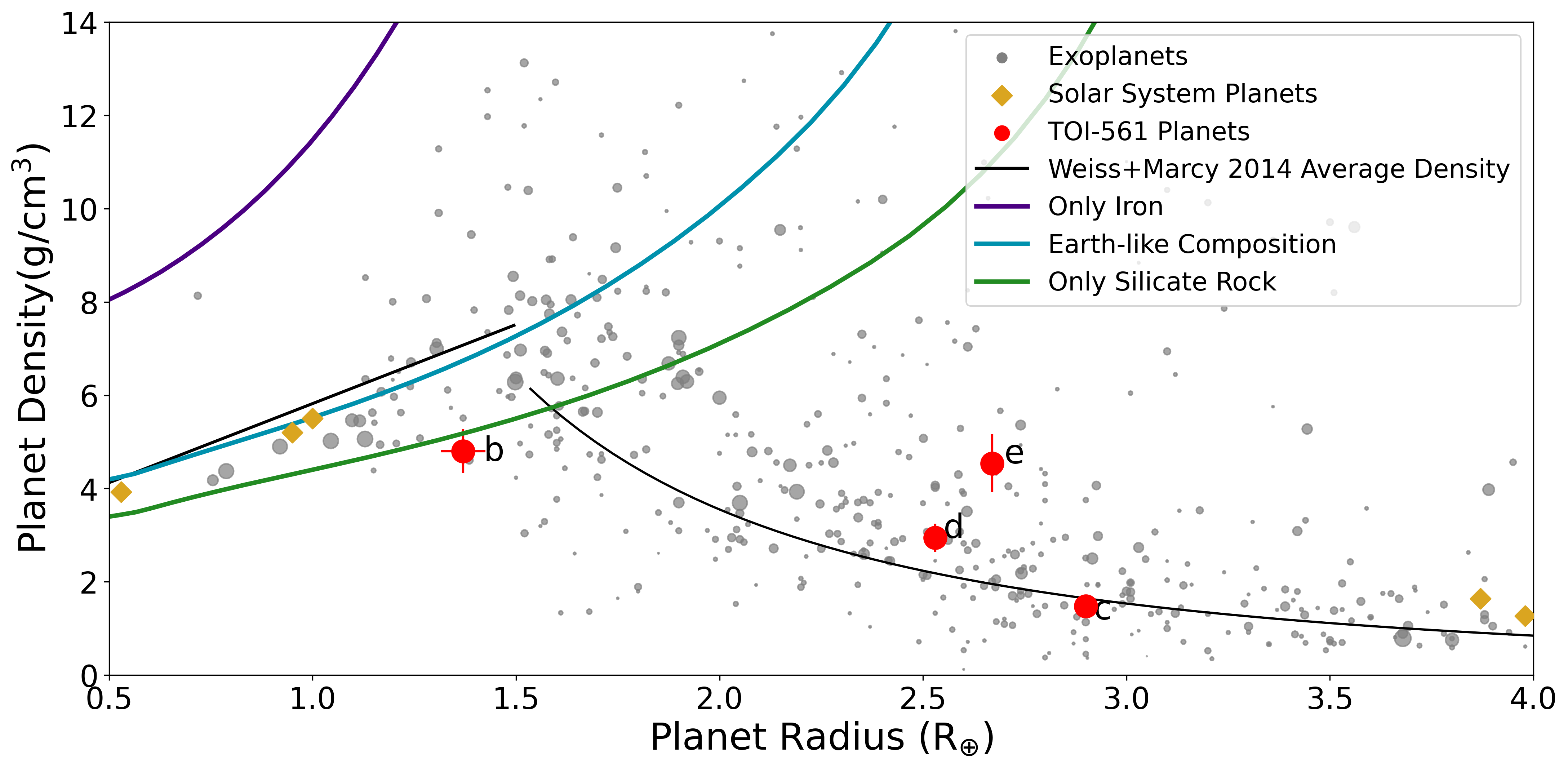}
    \caption{Planet mass (top panel) and density (bottom panel) as a function of radius for all known exoplanets with radii R$<$4R$_{\oplus}$ that have RV-determined mass measurements with fractional uncertainties $<$100$\%$ (grey points). The sizes of the points (excluding TOI-561 system planets) scale inversely with the fractional uncertainty of their mass measurement. The masses and densities for the planets orbiting TOI-561 are shown with their 1$\sigma$ uncertainties (red points).  The empirical mass-radius and density-radius relation from \citet{2014ApJ...783L...6W} are shown as black lines. The mass-radius and density-radius curves for planets of solid iron, solid rock, and an Earth-like composition from \cite{2019PNAS..116.9723Z} are shown  (colored lines).}
    \label{fig:density}
\end{figure*}

Figure \ref{fig:density} shows our mass and radius measurements for all four TOI-561 planets in the context of those from the broader sub-Neptune sized exoplanet population. We measured the density of TOI-561 b as $\rho_{b}$~=~\db ~$\pm$~ \dberr ~g/cm$^{3}$. This density places TOI-561 b as one of the lowest density planets with R$_{p} <$1.5R$_{\oplus}$\footnote{For perspective, a planet with an Earth-like composition at 1.37 R$_{\oplus}$ would have a density of 6.8 g/cm$^{3}$}. We also show the mass-radius and density-radius curves for planets of three different compositions that span the possible range for solid rocky planets: iron-only, Earth-like ($\sim$30$\%$ iron) and iron-free silicate rock \citep{2019PNAS..116.9723Z}. Our measurements for the mass and radius of TOI-561 b place it less than 1$\sigma$ below the curve for an iron-free silicate rock planet. This means that while TOI-561 b appears to be low density for a super-Earth sized planet, it is consistent both with solutions that include a gaseous envelope, and with those that require none. 

\section{Error Budget}

If TOI-561 b has a gaseous envelope, it would be one of the only planets in the ``rocky" planet regime ($R < 1.5 R_{/oplus}$) to host one \citep{2014ApJ...784...28K, 2018MNRAS.478..460A}. Before we proceed to composition modeling based on these mass and radius measurements, it is important to examine the uncertainties that restrict our ability to determine a precise density. 

\begin{figure*}
    \centering
    \includegraphics[width=1.0\textwidth]{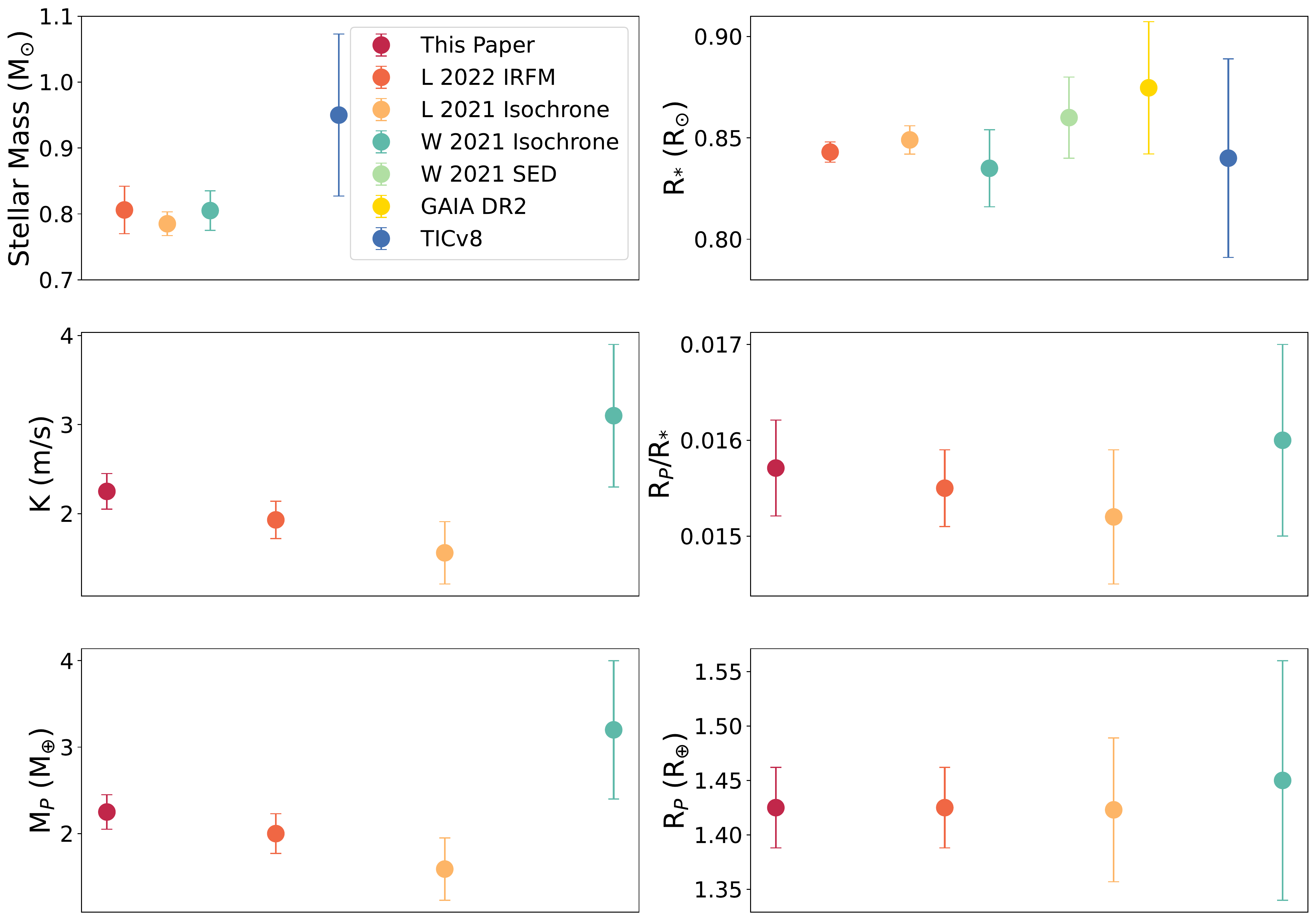}
    \caption{All available published values for the stellar mass and radius of TOI-561, along with semi-amplitude (K), R$_{p}$/R$_{*}$, planet mass (M$_{P}$), and planet radius (R$_{P}$) for TOI-561 b are shown with their 1$\sigma$ uncertainties. The values for stellar radius and mass shown here are found using  Infrared Flux Method (IRFM), SED fitting, or isochrone grid modeling. Legend Key: L 2022 = \cite{2022MNRAS.511.4551L}, L 2021 = \cite{2021MNRAS.501.4148L}, W 2021 = \cite{2021AJ....161...56W}.}
    \label{fig:stellar_params}
\end{figure*}

Our ability to characterize any exoplanet is limited by our ability to characterize its host star, as our measurements for the mass and radius of TOI-561 b are determined relative to the mass and radius of TOI-561. Estimates of stellar radius usually come from the Stefan-Boltzmann Law, 
\begin{equation}
    L_\odot=4\pi R_\odot^{2} \sigma_{\rm{SB}} T_{\rm{eff}}^{4}
\end{equation}
where $L_{\odot}$ is the luminosity, $R_{\odot}$ is the radius, $\sigma_{\rm{SB}}$ is the Stefan-Boltzmann constant, and $T_{\rm{eff}}$ is the effective temperature of the star. Measuring the luminosity of a star requires the distance to the star, along with bolometric flux. Assigning an uncertainty to bolometric flux involves propagating the uncertainties in atmospheric model grids, photometric zero-points, and reddening corrections. As a result, \cite{2022ApJ...927...31T} suggest a noise floor of $\sim$ 2$\%$ for measuring bolometric fluxes, and therefore luminosities. Effective temperatures of stars are defined through bolometric flux and angular diameter, and serve as fundamental calibrators for model-dependent methods such as high-resolution spectroscopy. This places a precision floor of $\sim$2$\%$ on effective temperature. Overall, \cite{2022ApJ...927...31T} recommend a noise floor of $\sim$4$\%$ in stellar radius, and $\sim$5$\%$ in stellar mass.

Angular diameter, bolometric flux, and effective temperature of stars are model parameters in the Infrared Flux Method (IRFM, \cite{1980A&A....82..249B}) or SED fitting \citep{2017AJ....153..136S}, which can then be used to infer stellar radii. Additionally, isochrone modeling can be used to infer the stellar radius (along with mass, density, and age) using measurements of effective temperature, metallicity, surface gravity, and parallax in conjunction with 3D dust maps \citep{2017ApJ...844..102H, 2016ApJ...823..102C}. 

Figure \ref{fig:stellar_params} shows the range of values for the mass and radius of TOI-561 across different publications and methods, along with their published uncertainties. \cite{2021MNRAS.501.4148L} use isochrones to measure stellar mass and radius, \cite{2022MNRAS.511.4551L} use the IRFM to determine a stellar radius, and \cite{2021AJ....161...56W} use both SED fitting and isochrone grid modeling. None of the values include systematic uncertainties to account for intrinsic uncertainties in measuring bolometric flux or effective temperature, and all four of these measurements fall above the 4$\%$ precision floor from \cite{2022ApJ...927...31T}. The scatter in stellar radius amongst all of the values in Figure \ref{fig:stellar_params} is $\sim$4.5$\%$, and the scatter of the four values discussed above is $\sim$3$\%$, which is in agreement with the \cite{2022ApJ...927...31T} prediction for method-dependent scatter. 

\begin{figure*}
    \centering
    \includegraphics[width=1.0\textwidth]{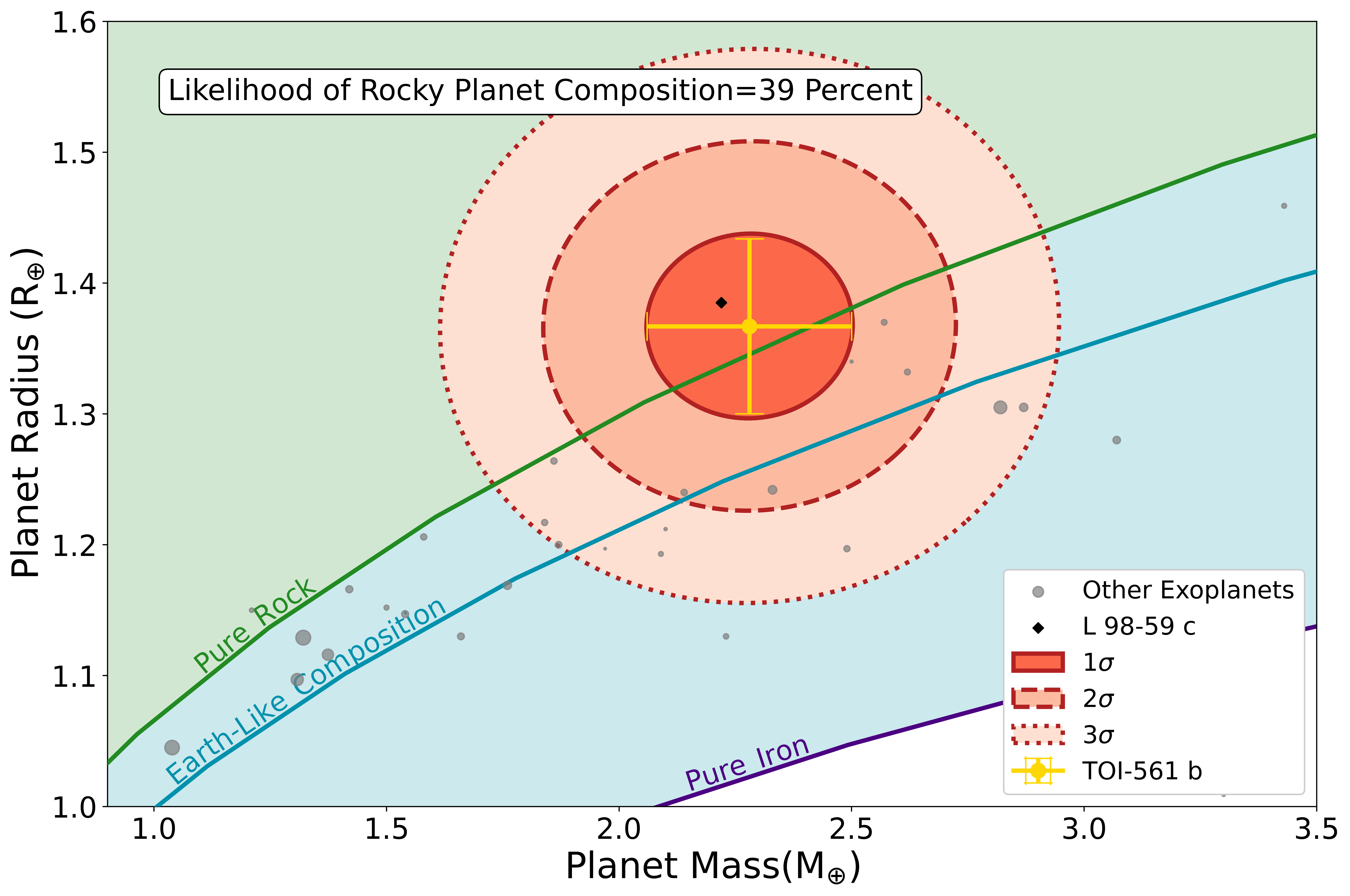}
    \caption{The mass and radius of TOI-561 b and uncertainties (1, 2, and 3 $\sigma$ contour regions) based on our new measurements (Table \ref{table:properties}). We used stellar parameters from \cite{2021AJ....161...56W}, and inflate the uncertainties on mass and radius to match the error floors established in \cite{2022ApJ...927...31T} (4$\%$ in radius, 5$\%$ in mass). The mass and radius of TOI-561 b are consistent with a rocky composition--a bulk composition containing only silicate rock and iron (blue shaded region)--in 40$\%$ of trials, and is consistent with a planet that needs a gaseous envelope (green shaded region) in 60$\%$ of trials. The population of known exoplanets with radius R$<$1.6R$_{\oplus}$ that have mass measurements are shown for context, and the size of the points (excluding TOI-561 b) scales inversely with the fractional uncertainty of their mass measurement. L98-59 c is highlighted as a recently characterized super-Earth with mass and radius measurements similar to those of TOI-561 b. }
    \label{fig:error}
\end{figure*}

To investigate the effects of these differences on planet composition, we constructed two sets of stellar parameters using the most conservative and the most optimistic cases. For the conservative case we use the isochrone radius and mass from \cite{2021AJ....161...56W}, but inflate the uncertainty in radius from 2.6$\%$ to 4$\%$. For the optimistic case we use the mass and radius from \cite{2022MNRAS.511.4551L}, with the 0.6$\%$ uncertainty on the radius as published. Figure \ref{fig:error} shows the 2-Dimensional mass and radius of distributions TOI-561 b found using the conservative cases. We drew 10000 samples of stellar mass and stellar radius from Gaussian distributions centered on the published values (although using the inflated $1\sigma$ error bars for our conservative case) \footnote{Stellar mass and radius are not independently varying parameters, so we tested the sensitivity of our results on the strength of the stellar mass-radius co-variance. We drew parameters from Gaussian distributions for stellar mass and radius assuming no co-variance, and we also drew assuming 100$\%$ co-variance and found no significant effect ($\sim$1$\%$ variation) in the final results due to the relatively larger uncertainties on semi-amplitude and transit depth. To ensure our errors are conservative where ever possible, we assume no co-variance in values reported.}. For each trial, we drew the transit depth and semi-amplitude from Gaussian distributions centered on the values measured in this work from Table \ref{table:properties}. We used these values to calculate the planet mass and radius in each trial, and recovered what fraction of our trials are consistent with a rocky planet solution and do not require a gaseous envelope or water layer. We deemed a planet as being consistent with a rocky planet composition if it falls below (i.e. more dense than) the``Pure Rock`` composition line in Figure \ref{fig:error} \citep{2019PNAS..116.9723Z}.

In the most conservative case ($4\%$ errors in stellar radius, using the isochrone-derived radius from \cite{2021AJ....161...56W}), we find that the mass and radius of TOI-561 b are consistent with a gas-free rocky composition $\sim$40$\%$ of the time (Figure \ref{fig:error}), while the optimistic case ($<1\%$ error in stellar radius) produces rocky planets only 21$\%$ of the time. Therefore, our measurements for planet mass and radius of TOI-561 b indicate that the planet is potentially too low density to have a rocky composition--however, this claim only holds with 1$\sigma$ significance with the most optimistic choice of stellar parameters, and less than 1$\sigma$ significance with a more conservative choice of stellar parameter uncertainties.

The parameters that contribute most to the uncertainty in planet composition are the transit depth and semi-amplitude, followed by the choice of stellar radius ($4\%$, Figure \ref{fig:stellar_params}). For the transit depth and semi-amplitude, additional measurements would help to shrink this uncertainty and would help determine the nature of this planet's composition. Meanwhile, our ability to precisely and accurately determine stellar radii will remain a limiting factor in our ability to characterize TOI-561 b.

\section{Planet Composition}
\label{sec:composition}

Given the uncertainty in planet parameters, we cannot determine if TOI-561 b is a rocky planet or has a gaseous envelope. We can, however, investigate the range of possible solid compositions given our planet mass measurement, and identify at what solid radius TOI-561 b would mirror the iron/rock building abundances present in the host star. We can investigate the effects of high melt fraction for the mantle, and briefly discuss possible envelope compositions.

Assuming negligible ice, water, or volatile fractions, we can express solid planet compositions in terms of the fraction of iron to total planet mass, or Core Mass Fraction (CMF). For this section, we adopt the stellar parameters found using isochrone models from \cite{2021AJ....161...56W} with errors inflated to 4$\%$ in stellar radius and 5$\%$ in stellar mass, and the RV semi-amplitude and R$_{P}$/R$_{*}$ listed in Table \ref{table:properties}. We performed interior modeling  using {\tt BurnMan 0.9} \citep{2016ascl.soft10010C}, which takes user-provided equations of state (EOS) and the masses of individual layers in a differentiated planet and computes the inner and outer radii of each layer. We incorporated {\tt BurnMan} in an iterative scheme which estimates the mass of a planet with a given radius, and specified mass distributions between the different layers (in this case, a solid metallic inner core, a liquid metallic outer core, a silicate mantle comprising MgSiO$_{3}$ (bridgmanite). Burnman does not incorporate atmospheric modeling, so our models account only for the solid portion of the planet. For a complete discussion on composition modeling using {\tt BurnMan}, see Brinkman et al. 2022 (in review).

We also used {\tt SuperEarth} to model the interior of TOI-561 b \citep{2006Icar..181..545V, 2007ApJ...656..545V}. Like {\tt BurnMan}, {\tt SuperEarth} divides planets into three primary layers (an iron core, rocky mantle, and ice/water layer) but instead of pure MgSiO$_{3}$, {\tt SuperEarth} uses a more nuanced mantle composition with four layers. The upper mantle includes olivene (MgSiO$_{4}$) and pyroxene, the transition zone features wadsleyite, ringwoodite, and pyroxene, the lower mantle includes bridgmanite and magnesiowustite, while the lower-most mantle has the same composition as the lower mantle, but features a high-pressure bridgmanite (post-perovskite). In addition, {\tt SuperEarth} takes user-provided mole fractions of silica inclusion in the iron core (here assumed to be 0), as well as iron mole fractions in the mantle (here assumed to be 0.1). A more thorough description of {\tt SuperEarth} can be found in \cite{2020MNRAS.499..932P}.

One source of uncertainty in our composition modeling is the degree to which the solid interior of TOI-561 b is differentiated. We assumed the planet is differentiated, a consequence of energy released and melting occurring during accretion \citep{2021ChEG...81l5735C}. However, the mass-radius relation of rocky planets is not very sensitive to their degree of differentiation  \citep{2008ApJ...688..628E}, with an expected difference of $\sim$2.5$\%$ in radius--smaller than our uncertainty on the radius itself \citep{2020MNRAS.499..932P}. 

To explore the range of possible compositions, we built planets that vary in composition from 0$\%$ iron core (CMF=0) to 100$\%$ iron (CMF=1.0) with a mass of \mb M$_{\oplus}$ and allow the radius to vary. This not only allows us to see what rocky planet compositions are possible within 1, 2, and 3, standard deviations from our measured radius, but also allows us to see the full range of interior solid compositions of various radii that could be present under a gaseous envelope. Our results using {\tt SuperEarth} are shown in Figure \ref{fig:comp}, but we note that the planet radii across all possible compositions held at fixed mass generated by {\tt BurnMan} and {\tt SuperEarth} differ by only $\sim$1$\%$. 

The smallest possible planet that produces a mass consistent with our mass measurement (although it does not match the radius measurement) is a core made of 100$\%$ iron, corresponding to a CMF of 1.0 at 1 R$_{\oplus}$. An iron core of this size would require a thick gaseous envelope to match the measured planet radius.\footnote{Note that Earth's atmosphere, which is one millionth of its mass, is optically thin and does not contribute to Earth's apparent radius, unlike the gaseous envelopes we consider.} The largest possible planet, made of 100$\%$ silicate rock corresponding to a CMF of 0, would have a radius of 1.35 R$_{\oplus}$. The best-fit radius (\radb R$_{\oplus}$) is larger than that of the largest solid rocky planet (CMF=0 planet), but is within 1$\sigma$ of it. Thus, our mass and radius measurements favor a silicate-rich, iron-poor planet (low CMF) if we assume the planet has no gas envelope, although the presence of a gas envelope would be consistent with any of the CMFs we tested. 

Melt fraction is another critical factor to consider in assessing the composition of TOI-561 b. The equilibrium temperature of the planet ($\sim$2300 K) indicates that a significant portion of the rocky mantle might be molten, which could increase the depth of the rocky mantle by up to 10$\%$ for a melt fraction of 1.0 \citep{2019A&A...631A.103B}. Without heat re-distribution from an atmosphere\footnote{We have no way with current measurements of knowing whether TOI-561 b has an either optically thin atmosphere or optically thick envelope made of high mean-molecular weight species. The melt fraction could therefore be larger than 0.5, which would put the radius of a planet with CMF=0.2 even closer to the measured radius of TOI-561 b.}, we would expect a melt fraction of $<$0.5, but this could still produce up to a 5$\%$ increase in rocky mantle depth, which would increase the total radius of the planet. Upper limits on radius size given a melt fraction of 0.5 are shown in grey on Figure \ref{fig:comp}. In the scenario in which TOI-561 b has a melt fraction close to 0.5, planet compositions with CMF=0, 0.1, and 0.2 are all consistent with our measured radius to within 1$\sigma$. 

\subsection{Host Star Composition}
The host star TOI-561 has abundance measurements for Silicon ([Si/H]=-0.24 $\pm$ 0.05 dex), Magnesium ([Mg/H]=-0.20 $\pm$ 0.05 dex), and Iron ([Fe/H]=-0.41 $\pm$ 0.05 dex) \citep{2021AJ....161...56W}, which allows us to compare Fe/Mg and Fe/Si ratios in the star to the expected Core-Mass Fraction of the planet given the protostellar nebular composition of the system. 

Using solar abundances from \cite{2019arXiv191200844L}, we find the absolute abundances (rather than relative to solar) of Mg, Si, and Fe in TOI-561. We then use the atomic weight of each species (55.8 u for Fe, 24.3 u for Mg, 28 for Si) to calculate the mass of each species relative to hydrogen. This gives us mass ratios of Fe/Si=1.25 $\pm$ 0.29, and Fe/Mg=1.18 $\pm$ 0.27.

\begin{figure*}
    \centering
    \includegraphics[width=1.0\textwidth]{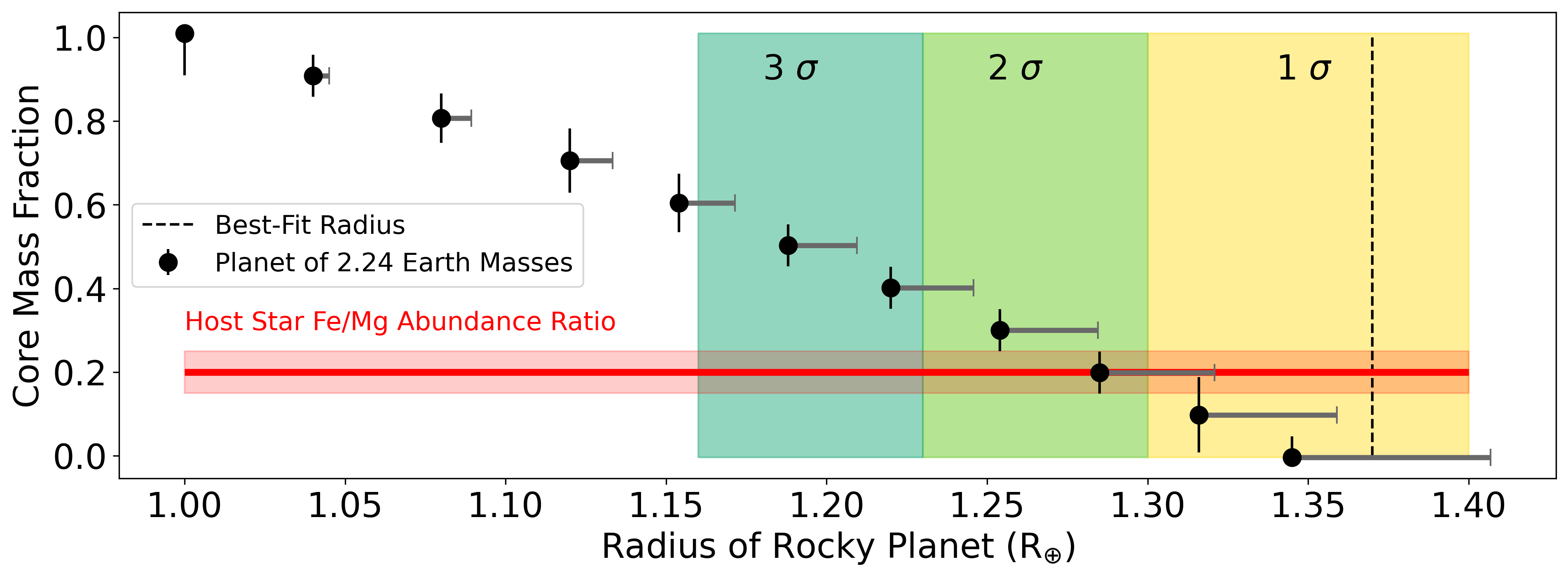}

    \caption{Here we show the range of possible compositions for the interior, rocky portion of TOI-561 b that are consistent within 1, 2, and 3 standard deviations of our measured planet radius and mass. We use {\tt Super-Earth} to simulate planets at varying radii, with varying fractions of iron and silicate rock that are expressed as the fraction of iron core mass to the total planet mass, also called Core Mass Fraction (CMF). The uncertainties in CMF represent the spread in compositions that produce a planet within 1$\sigma$ of the measured mass. The grey upper bound uncertainty in radius (grey) represents a $\sim$ 5 $\%$ increase in rocky mantle depth that planet would experience given a melt fraction in the mantle of 0.5 \citep{2019A&A...631A.103B}, inflating the total radius of the planet. The host star has an abundance ratio of Fe/Mg of 0.22, and the radius of a planet with a CMF=0.22 lies within 1 $\sigma$ of the measured radius of TOI-561 b given a high melt fraction.}
    \label{fig:comp}
\end{figure*}

Using the mantle mineral composition modeled using {\tt SuperEarth}, we can recover the expected mass ratio of Fe/Si and Fe/Mg given a specified Core Mass Fraction. A mass ratio of Fe/Si=1.25 $\pm$ 0.29 gives a CMF=0.22 $\pm$ 0.05. We find that a mass ratio of Fe/Mg=1.18 $\pm$ 0.27 is consistent with a CMF=0.20 $\pm$ 0.05. Magnesium is less volatile than Silicon and is thought to be better preserved through planet formation, therefore Fe/Mg is thought to be a better proxy for the pre-stellar nebula \citep{Yakovlev2018}. We have plotted the Fe/Mg derived CMF of 0.2 as the host star composition in Figure \ref{fig:comp}, but note that a CMF of either 0.20 or 0.22 for TOI-561 b produces a planet that is consistent to within 1$\sigma$ of our mass and radius measurements, assuming the rocky mantle has a high melt fraction.

The mass and radius of TOI-561 b are very similar to those of L 98-59 c \citep{2021A&A...653A..41D}, as seen in Figure \ref{fig:error}. L 98-59 c is also a super-Earth sized planet with mass and radius measurements that suggest it might be too low density for a solid rocky composition, but those measurements are also consistent with a rocky composition to within 1 $\sigma$. It is particularly interesting that while L 98-59 is an M Dwarf and TOI-561 is an early K dwarf, they are both metal poor ([Fe/H] $= -0.46 \pm 0.26$ and $-0.41 \pm 0.05$ dex, respectively). While further measurements are needed to measure the composition of either planet with a high degree of confidence, both of these systems are consistent with the hypothesis that metal-poor pre-stellar nebulae tend to form metal-poor rocky planets \citep{2018ApJ...867L...3B}.

\subsection{Possible Envelope Compositions}

While we cannot yet determine if TOI-561 b has an optically thick gaseous envelope, we can investigate some possible envelope compositions. Hydrogen-helium (H/He) atmospheres are thought to be the most common atmospheric composition for gaseous planets \citep{2020ApJ...891..111K, 2021JGRE..12606639B}. The measured masses and radii of exoplanets between radii between 1$<R_{\oplus}<$4 suggest that planets above 1.5$R_{\oplus}$ host these H/He envelopes (Mini-Neptunes) while planets below 1.5$R_{\oplus}$ do not (Super-Earths) \citep{2014ApJ...783L...6W, 2017AJ....154..109F}. Many of the planets that are currently smaller than 1.5R$_{\oplus}$ and have periods of less than 10 days are thought to have previously had H/He atmospheres during planet formation, like their Mini-Neptune cousins, which they subsequently lost \citep{, 2017ApJ...847...29O, 2019MNRAS.487...24G}. There are two stages during planet formation when this loss of atmosphere can occur: spontaneous mass loss following the dispersal of the planet-forming disk \citep{2018MNRAS.476..759G}, or, if the atmosphere is retained through disk dispersal, through subsequent core-powered mass loss \citep{2019MNRAS.487...24G} and/or photoevaporation \citep{2017ApJ...847...29O}.


In the case of spontaneous mass-loss, the pressure surrounding the planet suddenly drops when the disk disperses, leaving little to tie H/He envelopes to their host planet aside from gravity. \cite{2016ApJ...825...29G} give a criterion under which a planet will retain a H/He-dominated atmosphere during disk dispersal: 
\begin{equation}
\frac{M_{c}}{M_{\oplus}} > 6.3\left(\frac{T_{\rm{eq}}}{10^{3}K}\right)^{4/3}
\end{equation}
where $M_{c}$ is the mass of the rocky core of the planet and $T_{\rm{eq}}$ is the planet equilibrium temperature. For both sub-Neptunes and super-Earths, the majority of the planet's mass is contained within the rocky core \citep{2014ApJ...792....1L}, giving $M_{c}$=2.2 $M_{\oplus}$ for TOI-561 b. With an Earth-like albedo, TOI-561 b would have an equilibrium temperature T$_{eq}$=2300K. Under these conditions, TOI-561 b does not meet this criterion and would likely lose a H/He envelope spontaneously during disk dispersal. Letting albedo vary, we find that for any albedo $<$0.9995, TOI-561 b is too hot to retain an atmosphere through spontaneous mass loss. Given that the highest albedo object in the solar system is Enceladus at $\alpha$=0.81 \citep{2010Icar..206..573H}, it is highly unlikely that TOI-561 b has managed to retain a primordial H/He envelope. 




Several alternatives to H/He, particularly those with high mean molecular weight, are possible candidates for a gaseous envelope surrounding TOI-561 b.  \cite{2022MNRAS.511.4551L} suggest a liquid water layer or a water steam envelope. This scenario becomes more likely if the escape efficiency of H$_2$O atmospheres is much lower than for H$_2$-dominated atmospheres \citep{2022arXiv220706570Y}, or if very large quantities of H$_2$O are produced by oxidation of hydrogen by liquid magma \citep{2021ApJ...909L..22K}.  


Another species worth considering is CO$_{2}$. The escape efficiency for a pure-CO$_{2}$ atmosphere has been argued to be very low \citep{2009ApJ...703..905T}, so retaining a pure-CO$_{2}$ atmosphere would be easier than retaining the same mass of H$_{2}$O. However, even with low escape efficiency, the initial CO$_{2}$ content  needed would be greater than the (fractional) CO$_2$ content of Earth and Venus \citep{2009ApJ...703..905T,2020PNAS..11718264K}. Chondritic carbon values (carbon primarily in rock form as opposed to gaseous) for TOI-561 b would lead to a $\sim$1500 km (17$\%$ of total planet radius) CO$_{2}$/ CO+0.5O$_2$ atmosphere in equilibrium with a magma ocean, and be enough to explain its radius even when invoking an Earth-like interior (Peng $\&$ Valencia, in prep).

TOI-561 b is in also in the temperature regime where we would expect silicate rock to begin vaporizing \citep{2016ApJ...828...80K}, so a thin envelope could be produced from the evaporation of the rocky mantle. A silicate vapor atmosphere might be detectable in future through SiO absorption/emission \citep{2022A&A...661A.126Z}.

\section{Conclusion}
\label{sec:conclusion}

We collected 70 RVs using MAROON-X on Gemini and 42 RVs using HIRES on Keck I to improve the mass measurement of the ultra-short period super-Earth planet TOI-561 b. We combined our new RVs with literature RVs, literature photometry, and two new Sectors of TESS photometery to characterize the planet. Our main conclusions are as follows:
\begin{itemize}

\item We measure a mass of M$_{b}$=\mb ~$\pm$ \mberr ~$M_{\oplus}$, a radius of R$_{b}$=\radb~$\pm$~\radberr\ $R_{\oplus}$, and a density of $\rho_{b}$=~\db $\pm$ \dberr ~g/cm$^{3}$ for TOI-561 b. The mass, radius, and densities of all four planets in the TOI-561 system can be found in Table \ref{table:properties}.

\item While the low density of TOI-561 b suggests it might host a volatile envelope, it is consistent with an iron-poor rocky composition. Further, our measurements show that TOI-561 b is consistent within 1 $\sigma$ of being a rocky planet with an Core Mass Fraction of 0.2--matching the Fe/Si and Fe/Mg abundance ratios in its host star. TOI-561 b is consistent with the hypothesis that stars tend to form planets reflective of their abundance ratios.

\item If this planet indeed hosts a gas envelope, it is likely composed of high mean molecular weight species, differentiating it from the H/He envelopes that are typical of sub-Neptune sized planets. Envelope composition possibilities include those dominated by water or carbon dioxide, as well as evaporated silicates from the highly irradiated mantle.  

\item The largest sources of uncertainty on the density of TOI-561 b are the transit depth and RV semi-amplitude, followed by choice in stellar radius from literature values. We assessed the probability of TOI-561 b requiring a gaseous envelope by sampling the 2-Dimensional uncertainty space in planet mass and radius, and we find that the probability ranges from 60-80$\%$ depending on the choice of stellar parameters. 
\end{itemize}

TOI-561 b has the potential to be a rocky planet whose composition reflects the iron and rock-building element abundances in its host star--a common product of planet formation. It also has the potential to be a very unusual Super-Earth hosting a high-mean molecular weight envelope, potentially even made of evaporated rock. Until a more precise transit depth and a more accurate host star radius are determined, the interior and atmospheric composition of TOI-561 b will be difficult to constrain. Future observations with JWST may help us determine the presence of an atmosphere or gaseous envelope by measuring the day/night temperature differential and allowing us to infer heat transport across the surface of the planet.

\facilities{Transiting Exoplanet Survey Satellite (TESS), W. M. Keck Observatory, Gemini Observatory}

\software{Exoplanet \citep{exoplanet:zenodo}, Batman \citep{Kreidberg2015}, Lightcurve \citep{2018ascl.soft12013L}, RadVel \citep{2018PASP..130d4504F}, emcee \citep{2013PASP..125..306F}, BurnMan 0.9 \citep{2016ascl.soft10010C},SuperEarth \citep{2007ApJ...656..545V, 2020MNRAS.499..932P} NumPy \citep{harris2020array}, Matplotlib \citep{Hunter:2007}, pandas \citep{mckinney-proc-scipy-2010}, Astropy \citep{astropy:2013, astropy:2018, astropy:2022}, SciPy \citep{2020SciPy-NMeth} }

\section{Acknowledgements}

We recognize the cultural significance and sanctity that the summit of Mauna a Wākea has within the indigenous Hawaiian community. We are deeply grateful to have the opportunity to conduct observations from this mountain, while acknowledging the impact of our presence there and the ongoing efforts to preserve this special place in the universe.

C.L.B. is supported by the National Science Foundation Graduate Research Fellowship under Grant No. 1842402.

C.L.B., L.M.W. and D.H. acknowledge support from National Aeronautics and Space Administration (80NSSC19K0597) issued through the Astrophysics Data Analysis Program.

D.H. also acknowledges support from the Alfred P. Sloan Foundation.

C.L.B. and L.M.W. also acknowledge support from NASA Keck Key Stragetic Mission Support Grant No. 80NSSC19K1475.

M.B. is supported by the National Science Foundation Graduate Research Fellowship Grant No. DGE 1746045. 

J.M.A.M. is supported by the National Science Foundation Graduate Research Fellowship Program under Grant No. DGE-1842400. J.M.A.M. acknowledges the LSSTC Data Science Fellowship Program, which is funded by LSSTC, NSF Cybertraining Grant No. 1829740, the Brinson Foundation, and the Moore Foundation; his participation in the program has benefited this work.

We thank the time allocation committee of the University of Hawaii for supporting this work with observing time at the W. M. Keck Observatory and Gemini Observatory.

We gratefully acknowledge the efforts and dedication of the Keck Observatory and Gemini Observatory staff and representatives for observation support.

The development of the MAROON-X spectrograph was funded by the David and Lucile Packard Foundation, the Heising-Simons Foundation, the Gemini Observatory, and the University of Chicago. The MAROON-X team acknowledges support for this work from the NSF (award number 2108465) and NASA (through the \textit{TESS} Cycle 4 GI program, grant number 80NSSC22K0117). This work was enabled by observations made from the Gemini North telescope, located within the Maunakea Science Reserve and adjacent to the summit of Maunakea. We are grateful for the privilege of observing the Universe from a place that is unique in both its astronomical quality and its cultural significance.

\bibliography{toi561b.bib}
\bibliographystyle{aasjournal}

\end{document}